\documentclass[12pt,epsfig]{article}
\usepackage{float}
\usepackage{amsbsy}
\usepackage{amssymb}
\usepackage{color}

\newcommand{\beq}{\begin{equation}}
\newcommand{\eeq}{\end{equation}}
\newcommand{\bdm}{\begin{displaymath}}
\newcommand{\edm}{\end{displaymath}}
\newcommand{\beqr}{\begin{eqnarray}}
\newcommand{\eeqr}{\end{eqnarray}}
\newcommand{\beqrn}{\begin{eqnarray*}}
\newcommand{\eeqrn}{\end{eqnarray*}}

\def\bchi{\boldsymbol{\chi}}

\def\ve{\varepsilon}
\def\a{\alpha}
\def\k{\kappa}
\def\l{\lambda}

\def\bfm{{\bf m}}

\begin{document}

\title{On an approach for computing the generating functions  
of the characters of simple Lie algebras}

\author{Jos\'e Fern\'andez N\'u\~{n}ez$^{\dagger}$, Wifredo Garc\'{\i}a Fuertes$^{
\ddagger}$\\
\small Departamento de F\'\i sica, 
Facultad de Ciencias, Universidad de Oviedo, 33007-Oviedo, Spain\\
\small {\it $^\dagger$nonius@uniovi.es; $^\ddagger$wifredo@uniovi.es}\\
\and
 Askold M. Perelomov\\
\small Institute for Theoretical and Experimental Physics, 117259 Moscow, Russia.\\
\small {\it  aperelomof.uo@uniovi.es}}

\date{ }

\maketitle

\begin{abstract}\noindent
We describe a general approach to obtain the generating 
functions of the characters of simple Lie algebras which is based on the theory of the quantum trigonometric Calogero-Sutherland model. We show how the method works in practice by means of a few examples involving some low rank classical algebras. 
\end{abstract}
\bigskip


\medskip

\medskip

{\bf Short title:} On character generating functions of simple Lie algebras
\vfill\eject 
\section{Introduction}
The characters of the irreducible representations of the simple Lie algebras are systems 
of orthogonal polynomials which enjoy many interesting properties and have a distinguished role 
in pure mathematics and  mathematical physics \cite{mcdn}. Many features of orthogonal systems 
of this kind can be studied by means of their generating functions, defined as formal power 
series in some auxiliary variables whose coefficients give the polynomials entering in 
the system. In the case of the simple Lie algebras, after pioneering works such as \cite{bgw68} which deals with weight multiplicities or \cite{ps79} which refers to the characters themselves, several approaches for the computation of the generating functions have been proposed, see for instance \cite{ow07} and references therein. In general, these approaches are based on the Weyl character formula 
\cite{We25,We46} and require, in consequence, to perform in one way or other two quite cumbersome tasks: first, one has to sum over the elements of 
the Weyl group, which can be a rather involved combinatorial issue and, second, to finish with a Weyl-symmetric generating function, one has to divide two functions which are alternating with respect to Weyl reflections. 
While strategies of this type are powerful and can be used to establish formulas of great generality, 
see for instance \cite{ow07,fp02}, we feel that it would be desirable to put forward 
an alternative approach which, by avoiding these difficulties, be able to yield the generating functions of the characters of each particular 
algebra along a few simple steps. 

The main idea is to rely on the trigonometric Calogero-Sutherland model \cite{ca71}--\cite{op83}, rather than on the Weyl formula, as the tool for obtaining the characters. This point of view makes it possible to define a concrete and quite versatile procedure, which is suitable to be 
applied separately to each particular algebra, and uses Weyl symmetric instead of alternating formulas. The viability of such a procedure comes 
from two facts. First, in the last few years it has been shown explicitly how the quantum 
theory of integrable systems, in particular that 
of the Calogero-Sutherland model, can be used to compute 
the characters of the simple Lie algebras by identifying them with the eigenfunctions 
of the Hamiltonian for some particular values of the coupling constants, and several lists 
of characters obtained from this method are now available \cite{pe98a}--\cite{ffp03}. 
And, second, although the procedure that we are going to develop involves some long 
calculations,  they are quite straightforward and 
can be performed rather quickly by means of symbolic calculus languages like Mathematica 
or Maple, nowadays of common use. In this respect, we remark that, although we shall 
here illustrate the approach by applying it to some low-rank algebras, the use of 
these programs turns the method also useful for the higher-rank ones.

The paper is organized as follows. In the next section, after presenting a quick review of the theory of the quantum trigonometric Calogero-Sutherland systems, we deduce the differential equation which must satisfy the generating function of the characters and develop a method to solve it. In Section 3 we apply the procedure to some explicit examples, namely the classical  Lie algebras up to rank two. Section 4 offers some concluding comments. Finally, the results concerning the algebra $C_2$ needed in Section 3 are collected in the Appendix. All along the paper we make an extensive use of our previous results,  especially those contained in the last four articles in \cite{ffp03}, but both the approach described here and the results reported are completely new and were not the subject of the researches carried out in these references.

\section{Description of the method}
\subsection{Lie algebra characters and Calogero-Sutherland wave functions}
Let ${\cal A}$ be a simple Lie algebra of rank $r$ with simple roots $\alpha_1,\alpha_2,\ldots,\alpha_r$ and fundamental weights 
$\lambda_1,\lambda_2,\ldots,\lambda_r$. Let us denote $R_{\lambda}$ the irreducible 
representation of ${\cal A}$ with highest weight 
$\lambda=m_1 \lambda_1+m_2 \lambda_2+\cdots+m_r \lambda_r$. 
The character of this representation is defined as
\beq
\bchi_{m_1,m_2,\ldots,m_r}=\sum_{w} n_{w} e(w) \label{eq:zj}
\eeq
where the sum extends to all weights $w$ entering in the representation, $n_{w}$ is the 
multiplicity of the weight $w$ and, if $w=n_1 \lambda_1+n_2 \lambda_2+\cdots+n_r \lambda_r$, 
then  
\beq
e(w)=\exp\left(i \sum_{l=1}^r n_l \varphi_l\right)=x_1^{n_1} x_2^{n_2}\cdots x_r^{n_r}, 
\label{eq:ew}
\eeq
where $\varphi_1,\varphi_2,\ldots,\varphi_r$ are coordinates on the maximal torus and  
$x_l$ are complex phases, \mbox{$x_l=e^{i \varphi_l}$}.  

We can use the root system of ${\cal A}$ to define a very remarkable dynamical system, the trigonometric Calogero-Sutherland model. Here, we limit ourselves to mention the most salient features of this model which are useful for our present purposes and refer the reader to \cite{op83} for a more detailed treatment. The Hamiltonian has the form
\bdm
H=\frac{1}{2} p^2+ U(q)
\edm
where the coordinates $q=(q_1,q_2,\ldots,q_r)$ and momenta $p=(p_1,p_2,\ldots,p_r)$ are elements of a $r$-dimensional space $V$. The roots and weights of the algebra ${\cal A}$ can also be seen as elements of this same space, and the potential term is
\bdm
U(q)=\sum_{\a\in{\cal R}^+}\kappa_\a(\kappa_\a-1)\sin^{-2}\langle\a,q\rangle,
\edm
where ${\cal R}^+$ is the set of positive roots of ${\cal A}$  and $\langle\cdot,\cdot\rangle$ is the Euclidean scalar product on $V$. The constants $\kappa_\a$ must be chosen in such a way that the couplings $g_\a^2=\kappa_\a(\kappa_\a-1)$ are equal for roots of equal length. The model, which has many physical and mathematical applications \cite{dv00,po06}, is integrable both at the classical and quantum levels \cite{op76,op83}. In the latter case, it turns out that the energy eigenfunctions depend on $r$ quantum numbers ${\bf m}=(m_1,m_2,\ldots,m_r)$ and are of the form $\Psi_\bfm^\k=\Psi_0^\k\cdot\Phi_\bfm^\k$ where 
\bdm
\Psi_0^\k=\prod_{\alpha\in {\cal R}^+}\sin^{\kappa_\alpha}\langle\a,q\rangle
\edm
is the wave function of the ground state and the $\Phi_\bfm^\k$ are solutions of the related Schr\"odinger equation 
\beq
\Delta^\k\Phi_\bfm^\k=\ve(\bfm;\k)\,\Phi_\bfm^\k
\label{eq:sch}
\eeq
where $\Delta ^\k$ is the linear differential operator
\bdm
\Delta^\k=-\frac12\sum_{j=1}^r\partial_{q_j}^{\,2}-\sum_{\a\in {\cal
R}^+} \langle \a,\a\rangle \k_\a {\cot}\langle\a, q\rangle\,\langle\a,\partial_q\rangle
\edm
and the eigenvalues are 
\bdm
\ve(\bfm;\kappa)=2\langle\lambda+2\rho(\kappa),\lambda\rangle
\edm
for $2\rho(\kappa)=\sum_{\a\in{\cal R}^+}\k_\a\a$ and $\lambda$ the highest weight $\l=m_1\l_1+m_2\l_2+\dots+m_r\l_r$ defined by \bfm. The most relevant fact for us is that if we tune all coupling constants $\kappa_\alpha$ to one, the eigenfunctions of this      Schr\"{o}dinger operator are precisely the characters of the irreducible representations of the algebra \cite{op83}
\beq
\Phi_\bfm^1=\bchi_\bfm\,, \label{eq:phich}
\eeq
where the $\varphi$-angles are given in terms of the $q$-coordinates as $\varphi_j=2\langle \lambda_j,q\rangle$. 
This comes about as follows. Although the potential vanishes for
$\kappa_\alpha\to1$, there is a remnant of the interaction in that,
to take the limit consistently, we have to choose fermionic boundary
conditions ensuring that the wave functions are zero when
$\sin\langle\a,q\rangle=0$ for any positive root. As a consequence, the wave
function $\Psi_\bfm^1$ is given by a Weyl-alternating sum of free-particle
exponentials which turns out to coincide exactly with the numerator of the
Weyl character formula. The ground state wave function, on the other hand,
can be rewritten as the denominator of the Weyl formula, and the
$\Phi_\bfm^1$ are the characters of the Lie algebra thereby. (The particles
are free also for $\kappa_\alpha=0$, but in that case bosonic boundary
conditions are appropriate and the $\Phi_\bfm^0$ are the monomial symmetric
functions associated to the root system; the $\Phi_\bfm^\k$ for other values
of the couplings are systems of orthogonal polynomials which interpolate
between the monomial symmetric functions and the characters.)

Thus, we can obtain the characters by solving a second order differential equation. Furthermore, if we change variables and describe the dynamical system by means of the characters $z_k=\bchi_{\lambda_k}$ of the fundamental representations $R_{\lambda_k}$,  $k=1,2,\ldots,r$, the differential operator $\Delta^1$ takes the form
\beq
\Delta^1_z=\sum_{j, k=1}^ra_{jk}(z)\partial_{z_j}\partial_{z_k}+\sum_{j=1}^r
b_j(z)\partial_{z_j}, \label{eq:hamz}
\eeq
with $a_{jk}(z)$ and $b_j(z)$ polynomials in the $z_k$ with integer coefficients, and the Schr\"{o}dinger equation can be solved by iterative methods \cite{pe98a}--\cite{ffp03}. 
This operator can be given an explicit form taking into account that:
\begin{itemize}
\item $b_j(z)=\Delta_z^1z_j=\ve(0,\dots,1^{(j},\dots,0;1)z_j$,  and 
\item $\Delta^1_z(z_jz_k)=2a_{jk}(z)+b_j(z)z_k+b_k(z)z_j$,
\end{itemize} 
while $z_jz_k$ is the character of the tensor product $R_j\otimes R_k$. Hence, knowing all the quadratic Clebsch-Gordan series of the algebra we will be  able to determine the $a_{jk}(z)$ coefficients. For more explicit details, see \cite{ffp03} and the Appendix of this paper.

Once we know the definite expression for the operator $\Delta_z^1$,  it is possible to compute the characters $\bchi_\bfm$ as polynomials in the $z$-variables by solving the Schr\"odinger equation (\ref{eq:sch}) (with $\k=1$) in a recursive way  as follows. Given a weight $m_1\l_1+m_2\l_2+\cdots$, let us denote $z^{\bfm}=\Pi_{i=1}^rz_i^{m_i}$; thus the differential operator acting on $z^{\bfm}$ gives
\[
\Delta^1_zz^{\bfm}=\ve(\bfm;1)z^{\bfm}+\sum_{\beta\in\Lambda}R_{\bfm,\beta}\,z^{\bfm-\beta}\,,
\]
where $\Lambda$ includes only integral linear combinations of simple roots with positive coefficients. The polynomials $\bchi_\bfm$ can be written as $\bchi_\bfm(z)=z^{\bfm}+\sum_{\mu\in M}S_{\bfm,\mu}\,z^{\bfm-\mu}\,,$
where again only positive  powers of the $z$'s appear. By substituting in the Schr\"odinger equation (\ref{eq:sch}) we find the recursive formula for the coefficients of the polynomials,
\[
S_{\bfm,\mu}=\frac{1}{\ve(\bfm;1)-\ve(\bfm-\mu;1)}\,\sum_{\beta\in\Lambda}R_{\bfm-(\l-\beta),\beta}\,S_{\bfm,\mu-\beta}\,.
\]
This recursive formula is suitable for implementation in programs like Mathematica, Maple or others. For further details or an alternative method, see \cite{ffp03}.

\subsection{A differential equation for the generating function}
Our goal is to compute the generating function for the characters of ${\cal A}$, which is the formal series
\beq
G(t_j; z_k)=\sum_{m_1=0}^\infty\,\sum_{m_2=0}^\infty\cdots\sum_{m_r=0}^\infty t_1^{m_1} t_2^{m_2}
\cdots t_r^{m_r}\, \bchi_{m_1,m_2,\ldots,m_r}(z_1,z_2,\ldots,z_r)  \label{eq:gcar}
\eeq
in the auxiliary variables $t_1,t_2,\ldots,t_r$ and where we will treat the characters as polynomials in the $z$-variables. Now, from this definition and (\ref{eq:sch}) and (\ref{eq:phich}), we see that the generating function satisfies the differential equation
\beq
(\Delta_t-\Delta_z^1)\;G(t_j;z_k)=0\,,
\label{eq:test}
\eeq
where $\Delta_t$ is the differential operator
\bdm
\Delta_t=\ve(t_1\partial_{t_1},t_2\partial_{t_2},\ldots,t_r\partial_{t_r};1).
\edm
So, in principle, one could formulate the problem of finding the generating function as the problem of solving (\ref{eq:test}) with some suitable boundary conditions. Of course, posed in this abstract form, the problem seems to be difficult, but as we want the solution to be of the form (\ref{eq:gcar}), we can proceed in the reverse way, i.e. we will take (\ref{eq:gcar}) as the basis to build up a sensible ansatz and will afterward check that the ansatz satisfies (\ref{eq:test}).
\subsection{Solving the differential equation}
We expect that the solution of (\ref{eq:test}) which corresponds to the generating function is a rational function \cite{ps79}--\cite{We46}
\bdm
G(t_j;z_k)=\frac{N(t_j;z_k)}{D(t_j;z_k)}
\edm
where both the numerator $N(t_j;z_k)$ and the denominator $D(t_j;z_k)$ are polynomials 
in the $t$-variables with coefficients depending on the $z$-variables. We will find a  solution of this form through four successive steps. In the first three steps we will try to
build up $D(t_j;z_k)$ and $N(t_j;z_k)$ in a reasonable fashion. Then the fourth step is to check that the tentative form of $G(t_j;z_k)$ arising in this way does indeed satisfy the differential equation (\ref{eq:test}).
\begin{enumerate}
\item By Weyl invariance, the term $t_1^{m_1} t_2^{m_2}\cdots t_r^{m_r} \bchi_{m_1,m_2,\ldots,m_r}$ of the 
generating function includes a summand of the form 
$\prod_{j=1}^r e(w_j)^{m_j}$ for each possible combination formed by picking up a weight $w_j$ 
of each Weyl orbit $W\lambda_j$ of the fundamental weights. We can obtain all these products multiplied by 
$t_1^{m_1} t_2^{m_2}\cdots t_r^{m_r}$ if we multiply the geometric series 
$\sum_{p=0}^\infty (t_j)^p e(w)^p$ for all weights $w$ entering in the Weyl orbit 
of each $\lambda_j$. Therefore, the denominator of the generating function should be of the 
form
\beq
D=D_1\times D_2\times\cdots\times D_r \,,
\label{eq:d}
\eeq
where the factor $D_j$ is given by the formula
\beq
D_j=\prod_{w} (1-t_j e(w)) \label{eq:dj}
\eeq
with the product  extended to all 
the weights of $W\lambda_j$.\footnote{In fact, one can see that the use of the Weyl character formula leads to this form of $D$ as the common denominator in (\ref{eq:gcar}) \cite{ow07}.} Thus, $D_j$ is 
a polynomial in $t_j$ of degree equal to the cardinal $|\,W\lambda_j\,|$ of the orbit. The coefficients depend on the phases $x_l$ but, as the polynomial is by construction
invariant under Weyl reflections,  can be rewritten in terms of the Weyl invariant variables $z_k$. 
So, in this first step we compute each $D_j$ by means of (\ref{eq:dj}) and use (\ref{eq:zj}) 
and (\ref{eq:ew}) to put it as a function of the $z$-variables. Finally, (\ref{eq:d}) yields 
$D(t_j;z_k)$.
\item In the second step we compute the generating function
\beq
F(t_1,t_2,\ldots,t_r)=\sum_{m_1=0}^\infty\,\sum_{m_2=0}^\infty\cdots\sum_{m_r=0}^\infty t_1^{m_1} 
t_2^{m_2}\cdots t_r^{m_r}\dim R_{m_1 \lambda_1+m_2 \lambda_2+\cdots+m_r \lambda_r}  
\label{eq:f}
\eeq
for the dimensions of the irreducible representations of the algebra ${\cal A}$. This generating 
function is related to $G(t_j;z_k)$ through the replacement $x_l\to1$ or, equivalently, $z_k\rightarrow \dim R_{\lambda_k}$, 
i.e. $F(t_j)=G(t_j;\dim R_{\lambda_k})$.  The Weyl formula for dimensions implies that the coefficients of the series (\ref{eq:f}) are polynomials in the exponents $m_j$. Thus, $F(t_j)$ is a rational function which goes to zero for $t_j\to\infty$ at least as $t_j^{-1}$. So, we can write
\beq
F(t_j)=\frac{P(t_j)}{Q(t_j)} \label{eq:fpq}
\eeq
where the denominator is known,
\bdm
Q(t_j)=D(t_j;\dim R_{\lambda_k})=(1-t_1)^{\mid W\lambda_1\mid}(1-t_2)^{\mid W\lambda_2\mid}\cdots(1-t_r)^{\mid W\lambda_r\mid} ,
\edm
and the numerator should be of the form 
\beq
P(t_1,t_2,\ldots,t_r)=\sum_{l_1=0}^{\mid W\lambda_1\mid-1}\,\sum_{l_2=0}^{\mid W\lambda_2\mid-1}\cdots\sum_{l_r=0}^{\mid W\lambda_r\mid-1} t_1^{l_1} 
t_2^{l_2}\cdots t_r^{l_r}\, p_{l_1,l_2,\ldots,l_r}\,.
\label{eq:ansp}
\eeq
Here, $p_{0,0,\ldots,0}=1$ and the 
remaining numerical coefficients $p_{l_1,l_2,\ldots,l_r}$ are to be determined by comparing 
the Taylor expansion of (\ref{eq:fpq}) with the right member of (\ref{eq:f}). It will turn out 
that most of these coefficients are zero and the final form of $F(t_j)$ can be fixed after a few 
steps involving some very simple equations. 
\item In the third step we come back to the numerator $N(t_j;z_k)$ through the ansatz
\bdm
N(t_j;z_k)=\sum_{l_1,l_2,\ldots,l_r} C_{l_1,l_2,\ldots,l_r}(z_k)\, t_1^{l_1} t_2^{l_2}
\ldots t_r^{l_r}\,,
\edm
where, to simplify matters, we make the reasonable assumption that the only terms appearing in $N(t_j;z_k)$ are those corresponding to the nonvanishing $p$-coefficients. Often, we can guess some of the $C_{l_1,l_2,\ldots,l_r}(z_k)$ directly from the 
numerical value of the coefficient $p_{l_1,l_2,\ldots,l_r}$. For the others, we will have to 
write explicitly some low order terms of the right member of (\ref{eq:gcar}) and compare them 
with the Taylor expansion of $G(t_j;z_k)$. The  characters needed to accomplish this task  are polynomials in the $z$-variables which are easy to obtain by solving (\ref{eq:sch}) with $\Delta_z^1$ written as in (\ref{eq:hamz}). Once this is done, we have a tentative expression 
for $G(t_j;z_k)$.
\item The remaining step is to check that the simplifying assumption made in step 3 is valid and our tentative generating function is indeed correct. So, we substitute it in (\ref{eq:test}) and check that the differential equation is satisfied. Notice that, as we have manufactured our tentative generating function using only some low-order characters, this fourth step is necessary to prove that it gives (\ref{eq:gcar}) to all orders in the $t$-variables. 
\end{enumerate}

To sum up, the main ingredients needed to carry on the proposed procedure are: \emph{i)} the weights entering in the Weyl orbits and representations of the fundamental weights of the algebra, which can be obtained from sources like \cite{ov90}--\cite{lie}; \emph{ii)} the Weyl formula for dimensions \cite{ov90, otros}; and \emph{iii)} the Calogero-Sutherland Hamiltonian in $z$-variables and an iterative method for computing characters from it, see \cite{ffp03} for several examples. With this information and the help of some program for symbolic calculus, steps 1 to 4 can be carried out for any simple Lie algebra.
\section{Some examples}
Let us now apply the method described in the previous section to some Lie algebras in  order 
to appreciate how it works. We will work out the cases of the simple classical Lie algebras 
up to rank two, i.e. $A_1=B_1=C_1=D_1$, $A_2$ and $C_2\simeq B_2$, since $D_2\simeq A_1\oplus A_1$ is semisimple.

\subsection{The generating function for the algebra $A_1$}
This is the simplest case. The algebra has only one fundamental weight $\lambda_1$ and the 
Weyl orbit  $W\lambda_1$ has weights $\{\lambda_1,-\lambda_1\}$. The fundamental representation is $R_{\lambda_1}=W\lambda_1$, so the character $z_1$ is
\beq
z_1=e^{i \varphi_1}+e^{-i \varphi_1}=x_1+\frac{1}{x_1} \label{eq:a1z}
\eeq
The generating functions for characters and dimensions are
\bdm
G(t_1;z_1)=\sum_{m_1=0}^\infty t_1^{m_1} \bchi_{m_1}(z_1)\,,\qquad F(t_1)=
\sum_{m_1=0}^\infty t_1^{m_1} \dim R_{m_1} \,.
\edm
Let us now apply our four-step method. As there is only one weight, the denominator $D(t_1;z_1)$ 
is simply $D(t_1;z_1)=D_1$ with
\bdm
D_1=(1-t_1 x_1)(1-t_1\frac{1}{x_1})=1-t_1 z_1 +t_1^2 \,.
\edm
Now, as the dimension of $R_{\l_1}$ is two, the denominator of the generating function for 
dimensions is  $Q(t_1)=D(t_1;2)=(1-t_1)^2$. For the numerator, we try the ansatz 
\mbox{$P(t_1)=1+p_1 t_1$}. The Weyl formula for dimensions gives $\dim R_{m_1}=m_1+1$ 
and comparing this ansatz with the first few terms of the series $F(t_1)$, one readily finds 
that $p_1=0$. Therefore
\bdm
F(t_1)=\frac{1}{(1-t_1)^2} .
\edm
Given the form of $F(t_1)$, our ansatz for the numerator for the generating function for characters 
is
\bdm
N(t_1;z_1)=C_0(z_1)
\edm
but, given that $d_0=1$, the only sensible guess is $C_0(z_1)=1$ and the tentative expression 
for the generating function is
\beq
G(t_1;z_1)=\frac{1}{1-t_1 z_1+t_1^2} . \label{eq:gca1}
\eeq
It remains to confirm that this is correct. For that, we resort to the Calogero-Sutherland 
theory related to the algebra $A_1$ (see for example \cite{pe99}), which leads to the differential operator
\bdm
\Delta_z^1=(z_1^2-4)\partial_{z_1}^2+ 3 z_1 \partial_{z_1}\,,
\edm
with eigenvalues given by
\bdm
\varepsilon({m_1};1)=m_1^2+2 m_1 .
\edm
Thus, the operator $\Delta_t$ is $\Delta_t=t_1^2\partial^2_{t_1}+3 t_1 \partial_{t_1}$ and one 
can check that
\bdm
(\Delta_t-\Delta_z^1)\frac{1}{1-t_1 z_1+t_1^2}=0\,,
\edm
as it should be. Therefore (\ref{eq:gca1}) gives correctly the generating function for the 
characters of $A_1$.
\subsection{The generating function for the algebra $A_2$}
In this case there are two fundamental weights $\lambda_1, \lambda_2$, and the corresponding Weyl orbits $W\lambda_1$ and $W\lambda_2$ have, respectively, weights 
$\{\lambda_1,\lambda_2-\lambda_1,-\lambda_2\}$ and $\{\lambda_2,\lambda_1-\lambda_2,-\lambda_1\}$. The fundamental representations $R_{\lambda_1}$ and $R_{\lambda_2}$ have only one Weyl orbit. 
Therefore, the characters $z_1$ and $z_2$ are
\beqr
z_1&=&e^{i \varphi_1}+e^{-i \varphi_2}+e^{i (\varphi_2-\varphi_1)}=x_1+\frac{1}{x_2}+\frac{x_2}{x_1} 
\label{eq:a2z1}\\
z_2&=&e^{i \varphi_2}+e^{-i \varphi_1}+e^{i (\varphi_1-\varphi_2)}=x_2+\frac{1}{x_1}+
\frac{x_1}{x_2} \label{eq:a2z2}. 
\eeqr
The generating functions to be computed are
\beqr
G(t_1,t_2;z_1,z_2)&=&\sum_{m_1=0}^\infty \sum_{m_2=0}^\infty t_1^{m_1} t_2^{m_2}\,\bchi_{m_1,m_2}
(z_1,z_2)\label{eq:dgca2}\\
F(t_1,t_2)&=&\sum_{m_1=0}^\infty \sum_{m_2=0}^\infty t_1^{m_1} t_2^{m_2}\dim R_{m_1,m_2}
\label{eq:dgda2} 
\eeqr
and we follow the same steps than before. The denominator of the generating function for 
characters is $D(t_1,t_2;z_1,z_2)=D_1\times D_2$ where 
\beqrn
D_1&=&(1-t_1 x_1)(1-t_1\frac{1}{x_2})(1-t_1\frac{x_2}{x_1})=1-t_1 z_1+t_1^2 z_2-t_1^3\\
D_2&=&(1-t_2 x_2)(1-t_2\frac{1}{x_1})(1-t_2\frac{x_1}{x_2})=1-t_2 z_2+t_2^2 z_1-t_2^3.
\eeqrn
Since both $R_{\lambda_1}$ and $R_{\lambda_2}$ have dimension three, the denominator of the 
generating function for dimensions is $Q(t_1,t_2)=D(t_1,t_2;3,3)=(1-t_1)^3(1-t_2)^3$. For the 
numerator, we try the ansatz
\bdm
P(t_1,t_2)=\sum_{l_1=0}^2\, \sum_{l_2=0}^2 p_{l_1,l_2}\, t_1^{l_1} t_2^{l_2}
\edm
and, using that $\dim R_{m_1,m_2}=\frac{1}{2}(m_1+1)(m_2+1)(m_1+m_2+2)$ to compare with the 
right member of (\ref{eq:dgda2}), we find that the only nonvanishing coefficients are 
$p_{0,0}=-p_{1,1}=1$. Thus, the generating function for dimensions is
\bdm
F(t_1,t_2)=\frac{1-t_1 t_2}{(1-t_1)^3 (1-t_2)^3} .
\edm
In view of that, the ansatz for the numerator for the generating function for characters is
\bdm
N(t_1,t_2;z_1,z_2)=C_{0,0}(z_1,z_2)+C_{1,1}(z_1,z_2) t_1 t_2
\edm
and the reasonable guess is $C_{0,0}(z_1,z_2)=-C_{1,1}(z_1,z_2)=1$. One can confirm that this 
guess is correct by comparing the ansatz with the right member of (\ref{eq:dgca2}): to do so, 
the only characters needed  are
\bdm
\bchi_{1,0}(z_1,z_2)=z_1,\ \ \ \bchi_{0,1}(z_1,z_2)=z_2,\ \ \ \bchi_{1,1}(z_1,z_2)=z_1 z_2-1 .
\edm
 Therefore, the tentative form of the generating 
function is
\beq
G(t_1,t_2;z_1,z_2)=\frac{1-t_1 t_2}{(1-t_1 z_1+t_1^2 z_2-t_1^3)(1-t_2 z_2+t_2^2 z_1-t_2^3)} . 
\label{eq:gca2}
\eeq
Now, the second order differential operator associated to the Calogero-Sutherland model for 
the algebra $A_2$ whose eigenfunctions are the characters is \cite{pe98a, pe98b}
\bdm
\Delta_z^1=(z_1^2-3 z_2)\partial_{z_1}^2+(z_2^2-3 z_1)\partial_{z_2}^2+ (z_1 z_2-9)\partial_{z_1}
\partial_{z_2}+4 z_1 \partial_{z_1}+4 z_2 \partial_{z_2}
\edm
and their eigenvalues are
\bdm
\ve({m_1,m_2};1)=m_1^2+m_2^2+m_1 m_2+3 m_1 +3m_2.
\edm
Thus, the operator $\Delta_t$ is $\Delta_t=t_1^2\partial^2_{t_1}+t_2^2\partial^2_{t_2}+t_1 t_2 
\partial_{t_1} \partial_{t_2}+4 t_1 \partial_{t_1}+4 t_2 \partial_{t_2}$. One can check that, 
in fact,
\bdm
(\Delta_t-\Delta_z^1)\frac{1-t_1 t_2}{(1-t_1 z_1+t_1^2 z_2-t_1^3)(1-t_2 z_2+t_2^2 z_1-t_2^3)}=0
\edm
so that (\ref{eq:gca2}) is the correct generating function for the characters of $A_2$.

\subsection{The generating function for the algebra $C_2$}
Again, there are two weights, and the weights in the Weyl orbits 
$W\l_1$ and $W\l_2$ are, respectively, $\{\l_1,-\l_1,\l_1-
\l_2,\l_2-\l_1\}$ and $\{\l_2,-\l_2,2\l_1-\l_2,
-2\l_1+\l_2\}$. The fundamental representation $R_{\l_1}$ has only one Weyl orbit, but $R_{\l_2}$ includes also the null weight. Thus, the characters $z_1$ and $z_2$ are
\beqr
z_1&=&e^{i \varphi_1}+e^{-i \varphi_1}+e^{i(\varphi_1-\varphi_2)}+e^{i(\varphi_2-\varphi_1)}=x_1
+\frac{1}{x_1}+\frac{x_1}{x_2}+\frac{x_2}{x_1}\label{eq:b2z1}\\
z_2&=&1+e^{i \varphi_2}+e^{-i \varphi_2}+e^{i(2 \varphi_1-\varphi_2)}+e^{i(\varphi_2-2\varphi_1)}
=1+x_2+\frac{1}{x_2}+\frac{x_1^2}{x_2}+\frac{x_2}{x_1^2}\label{eq:b2z2}. 
\eeqr
The denominator of the generating function for characters is $D(t_1,t_2;z_1,z_2)=D_1\times D_2$
 with
\beqrn
D_1&=&(1-t_1 x_1)(1-t_1\frac{1}{x_1})(1-t_1\frac{x_1}{x_2})(1-t_1\frac{x_2}{x_1})=1-t_1 z_1
+t_1^2 (z_2+1)-t_1^3 z_1+t_1^4\\
D_2&=&(1-t_2 x_2)(1-t_2\frac{1}{x_2})(1-t_2\frac{x_1^2}{x_2})(1-t_2\frac{x_2}{x_1^2})
=1-t_2(z_2-1)+t_2^2(z_1^2-2z_2)-t_2^3( z_2-1)+t_2^4,
\eeqrn
while the denominator of the generating function for dimensions is 
\bdm
Q(t_1,t_2)=D(t_1,t_2;4,5)=(1-t_1)^4(1-t_2)^4.
\edm
For the numerator, we try the ansatz
\bdm
P(t_1,t_2)=\sum_{l_1=0}^3\, \sum_{l_2=0}^3 p_{l_1,l_2}\, t_1^{l_1} t_2^{l_2}
\edm
and, by means of $\dim R_{m_1,m_2}=\frac{1}{6}(m_1+1)(m_2+1)(m_1+m_2+2)(m_1+2m_2+3)$,  
we find that the only nonvanishing coefficients are $p_{0,0}=p_{0,1}=p_{2,1}=p_{2,2}=1$ and 
$p_{1,1}=-4$. Thus
\bdm
F(t_1,t_2)=\frac{(1+t_2)(1+t_1^2 t_2)-4 t_1 t_2}{(1-t_1)^4 (1-t_2)^4} .
\edm
Given this result, the ansatz for the numerator of the generating function for characters is
\bdm
N(t_1,t_2;z_1,z_2)=C_{0,0}(z_k)+C_{0,1}(z_k) t_2+C_{1,1}(z_k) t_1 t_2+C_{1,2}(z_k) t_1 t_2^2
+C_{2,2}(z_k) t_1^2 t_2^2
\edm
and the reasonable guess is $C_{0,0}=C_{0,1}=C_{2,1}=C_{2,2}=1$ and $C_{1,1}=-z_1$. Again, one 
can check that this guess is accurate by comparing with the explicit series of characters: 
the only characters needed are
$\bchi_{1,0},\bchi_{0,1},\bchi_{1,1},\bchi_{2,1}$ and $\bchi_{2,2}$, which are given in the 
Appendix. Therefore the tentative form of the generating function is
\beq
G(t_1,t_2;z_1,z_2)=\frac{1+t_2-z_1 t_1 t_2+t_1^2 t_2+t_1^2 t_2^2}{(1-(t_1+t_1^3) z_1+t_1^2 (z_2+1)+t_1^4)(1-(t_2+t_2^3)(z_2-1)+t_2^2(z_1^2-2z_2)+t_2^4)} . \label{eq:gcb2}
\eeq
Now we need the Calogero-Sutherland differential operator for $C_2$, which is shown in the 
Appendix to be
\bdm
\Delta_z^1=(z_1^2-2 z_2-6)\partial_{z_1}^2+(2z_2^2-4 z_1^2+4 z_2-6)\partial_{z_2}^2
+ (2z_1 z_2-10 z_1)\partial_{z_1}\partial_{z_2}+5 z_1 \partial_{z_1}+4 z_2 \partial_{z_2} .
\edm
Since their eigenvalues are
\bdm
\varepsilon({m_1,m_2};1)=m_1^2+2 m_2^2+2 m_1 m_2+4 m_1 +6m_2,
\edm
we see that the operator $\Delta_t$ is $\Delta_t=t_1^2\partial^2_{t_1}+2 t_2^2\partial^2_{t_2}
+2 t_1 t_2 \partial_{t_1} \partial_{t_2}+5 t_1 \partial_{t_1}+8 t_2 \partial_{t_2}$. We now can 
check that
\bdm
(\Delta_t-\Delta_z^1)\frac{1+t_2-z_1 t_1 t_2+t_1^2 t_2+t_1^2 t_2^2}{(1-(t_1+t_1^3) z_1+t_1^2 (z_2+1)+t_1^4)(1-(t_2+t_2^3)(z_2-1)+t_2^2(z_1^2-2z_2)+t_2^4)}=0
\edm
and this proves that (\ref{eq:gcb2}) is the correct generating function for the characters 
of $C_2$.

\subsection{Generating functions for some subsets of characters}
If in the previous examples for $A_2$ and $C_2$ we take $t_1=0$ or $t_2=0$, we obtain the generating function for the characters of the form $\bchi_{m_1,0}(z_1,z_2)$ or $\bchi_{0,m_2}(z_1,z_2)$ of these algebras. It is also possible to use the method that we have been describing
 to compute the generating function of some other subsets of characters. Let us, for instance, 
consider the case of the diagonal characters of $A_2$ and seek for their generating function 
\bdm
G_{diag}(t;z_1,z_2)=\sum_{m=0}^\infty t^{m} \bchi_{m,m}(z_1,z_2).
\edm
Now, according to the reasoning of Sect. 2, the denominator should be a product including 
all factors of the form $(1-t\, e(w_1) e(w_2))$ for all couples of nonzero weights $w_1$ 
and $w_2$ entering in the Weyl orbits $W\lambda_1$ and $W\lambda_2$ of $A_2$. 
This gives
\bdm
D_{diag}(t;z_1,z_2)=1 + t^6 + (3 - z_1z_2)(t+t^5)+ 
    (6 + z_1^3 - 5z_1z_2 + z_2^3)(t^2+t^4) + 
    (7 + 2z_1^3 - 6z_1z_2 - z_1^2z_2^2 + 2z_2^3)t^3 .
\edm
From this, we can find easily the generating function for the dimensions of the diagonal 
characters, which is
\bdm
F_{diag}(t)=\frac{1+2t-6t^2+2t^3+t^4}{(1-t)^6}
\edm
Thus, we try the ansatz
\bdm
N_{diag}(t;z_1,z_2)=\sum_{l=0}^4 C_l(z_1,z_2)\,t^l
\edm
for the numerator, and using the explicit form of the characters $\bchi_{m,m}(z_1,z_2)$ 
for $m=0$ to 4, we arrive to a  tentative form of $G_{diag}(t;z_1,z_2)$ as
\beq
G_{diag}(t;z_1,z_2)=\frac{1 + 2t -(z_1z_2-3)t^2+ 2t^3 + t^4}{
  D_{diag}(t;z_1,z_2)} \label{eq:diag}
\eeq
For this subset of characters, the operator $\Delta_t$ takes the form $\Delta_t=3 t^2 
\partial_t^2+9 t \partial_t$ and one can check that $(\Delta_t-\Delta_z^1)G_{diag}(t;z_1,z_2)=0$. 
Thus, (\ref{eq:diag}) gives in fact the correct generating function.

The same procedure applied to $C_2$ gives the generating function of diagonal characters  of that algebra. The result is
\beqrn
G_{diag}(t;z_1,z_2)=\frac{(1 - t^2)(1 + t^4 + 2tz_1 + 2t^3z_1 + t^2(2z_1^2 -z_1^2z_2 + z_2 + z_2^2))}
 {1 + t^8 - (t+t^7) d_1+ (t^2+t^6)d_2 +    (t^3+t^5) d_3 +   t^4 d_4}
\eeqrn
where
\beqrn
d_1&=&z_1(-3 + z_2)\\
d_2&=&-1 + z_1^4 + z_1^2(3 - 6z_2) + z_2 + 3z_2^2 + z_2^3\\
d_3&=&z_1(-3 + 2z_1^4 - 2z_2 + 8z_2^2 + 3z_2^3 - 
         z_1^2(-2 + 9z_2 + z_2^2))\\
d_4&=&z_1^6 + z_1^4(4 - 6z_2) + z_1^2(-5 - 6z_2 + 5z_2^2) + 
         z_2(-2 + 3z_2 + 4z_2^2 + z_2^3).
\eeqrn
\section{Concluding remarks}
We have presented an approach for the computation of the generating function of the characters of a simple Lie algebra which is based on the theory of an integrable mechanical system, namely the quantum trigonometric Calogero-Sutherland model. The key point is that the Schr\"{o}dinger equation of that model leads to a differential equation for the generating function. This equation can be solved by means of a convenient ansatz which has been described in detail. The procedure involves the computation of some low order characters and this can be done by solving the Calogero-Sutherland Schr\"{o}dinger equation.  The approach is by design Weyl invariant. We avoid the use of alternating functions by formulating all computations in terms of a set of dynamical variables $z$ which correspond to the characters of the fundamental representations of the algebra. We have illustrated our approach by applying it to some low rank classical algebras, but we expect it to be equally useful for the classical or exceptional higher rank ones. Of course, in these cases the calculations are longer, but the type of mathematical objects that they involve are polynomials in $z_k$ variables with integer coefficients. This make the computations especially well suited for the use of programs like Mathematica or Maple. In fact, we find it likely that this approach is more efficient than others based on the Weyl character formula. At any event, when one is faced to some complicated problem as it is the computation of the generating functions for characters in closed form, it is always desirable to have the possibility of choosing among different approaches.

There are other interesting mathematical objects, such as the Weyl invariant monomial functions of some zonal spherical functions in symmetric spaces, which are also identical to the eigenfunctions of the Calogero-Sutherland model for adequate values of the coupling constants. It could be that an approach similar to that advocated for in this paper be useful for obtaining generating functions for them.

Finally, let us take the example of $A_2$ to comment on an alternative way to obtain the denominator $D(t;z)$ which could be useful for higher rank algebras. If we take $t_2=0$ in (\ref{eq:gca2}) we have:
\bdm
D_1\, G(t_1,0;z_1,z_2)=(1-t_1 z_1+t_1^2 z_2-t_1^3) \sum_{m_1=0}^\infty t_1^{m_1} \bchi_{m_1,0}(z_1,z_2)=1
\edm
and this implies the recurrence relation
\beq
\bchi_{m_1,0}-z_1 \bchi_{m_1-1,0}+z_2 \bchi_{m_1-2,0}-\bchi_{m_1-3,0}=0\quad {\rm for}\  m_1 > 3 .\label{eq:recu}
\eeq
This recurrence relation can also be easily obtained by combining two well-known Clebsch-Gordan series of $A_2$, see for instance \cite{pe98a}
\beqrn
z_1 \bchi_{m_1,0}&=&\bchi_{m_1+1,0}+\bchi_{m_1-1,1}\\
z_2 \bchi_{m_1,0}&=&\bchi_{m_1,1}+\bchi_{m_1-1,0} .
\eeqrn
Thus, by working backwards, we could have obtained $D_1$ from the Clebsch-Gordan series. This has the advantage that the result appears directly written in the $z$ variables. For an arbitrary simple algebra, $D_j$ is of degree $|\,W\lambda_j\,|$ and the characters  $\bchi_{0,\ldots,0,m_j,0,\ldots,0}$ for $m_j >|\, W\lambda_j\,|$ will obey a recurrence relation of $|\, W\lambda_j\,|+1$ terms similar to (\ref{eq:recu}). Now, if we know the Clebsch-Gordan series of type $z_k \bchi_\bfm$ for the algebra, we can combine them to obtain this recurrence relation and, in this way, we can compute the factors $D_j$ appearing in the denominator of the generating function. Some useful sources for Clebsch-Gordan series are \cite{ov90}--\cite{lie} and for examples of how to compute the Clebsch-Gordan series using the Calogero-Sutherland Schr\"{o}dinger equation see \cite{pe98a} and the second and third papers in \cite{ffp03}.
\section*{Appendix}
In this Appendix, we present the Calogero-Sutherland differential operator whose eigenfunctions 
are the characters of the algebra $C_2$ and give a short collection of characters of $C_2$ 
which includes those needed to fix the numerator of the generating function 
$G(t_1,t_2;z_1,z_2)$. For a more detailed account of the use of the quantum Calogero-Sutherland 
models to deal with the characters or other generalized orthogonal polynomials related to Lie algebras, see \cite{ffp03} and references therein.

The general form of the differential operator we are looking for is 
\bdm
\Delta_z^1=a_{1,0}(z_1,z_2)\partial_{z_1}^2+a_{0,1}(z_1,z_2)\partial_{z_2}^2+a_{1,1}(z_1,z_2)
\partial_{z_1}\partial_{z_2}+b_1(z_1,z_2)\partial_{z_1}+b_2(z_1,z_2)\partial_{z_2}
\edm
and we will fix the coefficients using that the character $\bchi_{m_1,m_2}$ of the irreducible 
representation of $C_2$ with highest weight $\l=m_1\l_1+m_2\l_2$ is 
an eigenfunction of the Schr\"{o}dinger equation 
\bdm
\Delta_z^1\bchi_{m_1,m_2}=\varepsilon(m_1,m_2;1)\,\bchi_{m_1,m_2}
\edm
with eigenvalue 
\bdm
\varepsilon(m_1,m_2;1)=\langle\l,\l\rangle+2\langle\l,\rho\rangle,
\edm
where $\langle\cdot,\cdot\rangle$ is the scalar product in the space spanned by the orthonormal 
basis $\{e_1,e_2\}$, the two fundamental weights of $C_2$ are represented in this basis by
\bdm
\l_1=e_1,\hspace{2cm}\l_2=e_1+e_2 ,
\edm
the four positive roots are
\bdm
\a_1=e_1-e_2,\hspace{1cm}\a_2=2e_2,\hspace{1cm}\a_3=e_1+e_2,\hspace{1cm}\a_4=2e_1,
\edm
and $\rho=\l_1+\l_2$ is the Weyl vector of the algebra \cite{ov90,otros}, so that
\bdm
\varepsilon({m_1,m_2};1)=m_1^2+2 m_2^2+2 m_1 m_2+4 m_1 +6m_2 .
\edm
On the other hand, given that 
\beqrn
\bchi_{1,0}(z_1,z_2)&=&z_1\\
\bchi_{0,1}(z_1,z_2)&=&z_2 ,
\eeqrn
the direct products of the representations of $C_2$ \cite{ov90, otros}
\beqrn
R_{1,0}\otimes R_{1,0}&=&R_{2,0}+R_{0,1}+R_{0,0}\\
R_{1,0}\otimes R_{0,1}&=&R_{1,1}+R_{1,0}\\
R_{0,1}\otimes R_{0,1}&=&R_{0,2}+R_{2,0}+R_{0,0}
\eeqrn
allow us to solve for
\beqrn
\bchi_{2,0}(z_1,z_2)&=&z_1^2-z_2-1\\
\bchi_{1,1}(z_1,z_2)&=&z_1 z_2-z_1\\
\bchi_{0,2}(z_1,z_2)&=&z_2^2-z_1^2+z_2 .
\eeqrn
Using all these characters in the Schr\"{o}dinger equation one finds
\beqrn
a_{1,0}(z_1,z_2)&=&z_1^2-2 z_2-6\\
a_{0,1}(z_1,z_2)&=&2z_2^2-4 z_1^2+4 z_2-6\\
a_{1,1}(z_1,z_2)&=&2z_1 z_2-10 z_1\\
b_1(z_1,z_2)&=&5 z_1\\
b_(z_1,z_2)&=&4 z_2 .
\eeqrn
Once the operator $\Delta_z^1$ is known, other characters of $C_2$ can be computed by solving 
the Schr\"{o}dinger equation. 
We give here a few of them:
\beqrn
\bchi_{0,3}(z_1,z_2)&=&-1 + z_1^2 - 2 z_1^2 z_2 + 2 z_2^2 + z_2^3 \\
\bchi_{1,2}(z_1,z_2)&=&z_1 - z_1^3 + z_1 z_2^2 \\
\bchi_{2,1}(z_1,z_2)&=&1 - z_1^2 - z_2 + z_1^2 z_2 - z_2^2 \\
\bchi_{3,0}(z_1,z_2)&=&-z_1 + z_1^3 - 2 z_1 z_2 \\
\bchi_{0,4}(z_1,z_2)&=&-z_1^2 + z_1^4 - 2 z_2 + z_2^2 - 3 z_1^2 z_2^2 + 
   3 z_2^3 + z_2^4 \\
\bchi_{1,3}(z_1,z_2)&=&-2 z_1 + 2 z_1^3 - 2 z_1^3 z_2 + z_1 z_2^2 + 
   z_1 z_2^3 \\
\bchi_{2,2}(z_1,z_2)&=&2 z_1^2 - z_1^4 + z_1^2 z_2 - 2 z_2^2 + z_1^2 z_2^2 - 
   z_2^3 \\
\bchi_{3,1}(z_1,z_2)&=&2 z_1 - z_1^3 + z_1^3 z_2 - 2 z_1 z_2^2\\ 
\bchi_{4,0}(z_1,z_2)&=&-z_1^2 + z_1^4 + 2 z_2 - 3 z_1^2 z_2 + z_2^2 \\
\bchi_{0,5}(z_1,z_2)&=&3 z_1^2 - 2 z_1^4 - 2 z_2 + 3 z_1^4 z_2 - 3 z_2^2 - 
   3 z_1^2 z_2^2 + 3 z_2^3 - 4 z_1^2 z_2^3 + 4 z_2^4 + z_2^5 \\
\bchi_{1,4}(z_1,z_2)&=&2 z_1 - 3 z_1^3 + z_1^5 - 2 z_1 z_2 + 2 z_1^3 z_2 - 
   3 z_1^3 z_2^2 + 2 z_1 z_2^3 + z_1 z_2^4 \\
\bchi_{2,3}(z_1,z_2)&=&1 - 4 z_1^2 + 2 z_1^4 + 2 z_2 + z_1^2 z_2 - 
   2 z_1^4 z_2 - z_2^2 + 3 z_1^2 z_2^2 - 3 z_2^3 + z_1^2 z_2^3 - z_2^4 \\ 
\bchi_{3,2}(z_1,z_2)&=&-z_1 + 2 z_1^3 - z_1^5 + 2 z_1^3 z_2 - 2 z_1 z_2^2 + 
   z_1^3 z_2^2 - 2 z_1 z_2^3 \\
\bchi_{4,1}(z_1,z_2)&=&-1 + 2 z_1^2 - z_1^4 - z_2 + z_1^2 z_2 + z_1^4 z_2 + 
   2 z_2^2 - 3 z_1^2 z_2^2 + z_2^3 \\
\bchi_{5,0}(z_1,z_2)&=&-z_1 - z_1^3 + z_1^5 + 4 z_1 z_2 - 4 z_1^3 z_2 + 
   3 z_1 z_2^2 \\
\bchi_{0,6}(z_1,z_2)&=&1 - 3 z_1^2 + 3 z_1^4 - z_1^6 + 6 z_1^2 z_2 - 
   3 z_1^4 z_2 - 6 z_2^2 + 6 z_1^4 z_2^2 - 3 z_2^3 - 8 z_1^2 z_2^3 + 
   6 z_2^4 - 5 z_1^2 z_2^4 \\&+& 5 z_2^5 + z_2^6 \\
\bchi_{1,5}(z_1,z_2)&=&-2 z_1 + 6 z_1^3 - 3 z_1^5 - 2 z_1^3 z_2 + 
   3 z_1^5 z_2 - 3 z_1 z_2^2 + z_1 z_2^3 - 4 z_1^3 z_2^3 + 3 z_1 z_2^4 + 
   z_1 z_2^5 \\
\bchi_{2,4}(z_1,z_2)&=&-1 + 4 z_1^2 - 4 z_1^4 + z_1^6 + 2 z_2 - 
   3 z_1^2 z_2 + z_1^4 z_2 + 3 z_2^2 + 3 z_1^2 z_2^2 - 3 z_1^4 z_2^2 - 
   3 z_2^3 \\&+& 5 z_1^2 z_2^3 - 4 z_2^4 + z_1^2 z_2^4 - z_2^5\\
\bchi_{3,3}(z_1,z_2)&=&2 z_1 - 5 z_1^3 + 2 z_1^5 + 4 z_1 z_2 - z_1^3 z_2 - 
   2 z_1^5 z_2 + 5 z_1^3 z_2^2 - 4 z_1 z_2^3 + z_1^3 z_2^3 - 2 z_1 z_2^4  \\
\bchi_{4,2}(z_1,z_2)&=&-z_1^2 + 2 z_1^4 - z_1^6 - z_2 - 3 z_1^2 z_2 + 
   3 z_1^4 z_2 + z_2^2 - 3 z_1^2 z_2^2 + z_1^4 z_2^2 + 3 z_2^3 - 
   3 z_1^2 z_2^3 + z_2^4 \\
\bchi_{5,1}(z_1,z_2)&=&-z_1 + 2 z_1^3 - z_1^5 - 5 z_1 z_2 + 2 z_1^3 z_2 + 
   z_1^5 z_2 + 3 z_1 z_2^2 - 4 z_1^3 z_2^2 + 3 z_1 z_2^3 \\
\bchi_{6,0}(z_1,z_2)&=&1 - 2 z_1^2 - z_1^4 + z_1^6 - z_2 + 6 z_1^2 z_2 - 
   5 z_1^4 z_2 - 3 z_2^2 + 6 z_1^2 z_2^2 - z_2^3\,.
\eeqrn

\section*{Acknowledgement}

J.F.N. acknowledges financial support from MTM2012-33575 project, SGPI-DGICT(MEC), Spain.

\end{document}